\documentclass[%
 aip,
 amsmath,amssymb,
 reprint,%
]{revtex4-1}

\usepackage{graphicx}
\usepackage{dcolumn}
\usepackage{bm}
\usepackage[T1]{fontenc}
\usepackage[utf8]{inputenc}
\usepackage{mathptmx}
\usepackage{etoolbox}
\usepackage{sidecap}
\usepackage{amsfonts}
\usepackage{mathtools,halloweenmath}
\usepackage{array}
\DeclareUnicodeCharacter{148}{\snowball}

\makeatletter
\def\@email#1#2{%
 \endgroup
 \patchcmd{\titleblock@produce}
  {\frontmatter@RRAPformat}
  {\frontmatter@RRAPformat{\produce@RRAP{*#1\href{mailto:#2}{#2}}}\frontmatter@RRAPformat}
  {}{}
}%
\makeatother
\begin{document}

\preprint{AIP/123-QED}

\title[Investigating the classical problem of pursuit, in two modes]{Investigating the classical problem of pursuit, in two modes}
\author{A.H. Arshadi Kalameh}
\affiliation{Department of Physics, Isfahan University of Technology, Isfahan, Iran 
}%

\author{K. Bayati Komitaki}%
\affiliation{Department of Physics, Isfahan University of Technology, Isfahan, Iran 
}%

\author{R. Sharifian}
\affiliation{%
Department of Physics, Sharif University of Technology, Tehran, Iran
}%

\author{M.M. Eftekhari}
\affiliation{Department of Physics, Isfahan University of Technology, Isfahan, Iran 
}%

\date{4 September 2023}

\begin{abstract}
The pursuit problem is a historical issue of the application of mathematics in physics, which has been discussed for centuries since the time of Leonardo Da Vinci, and its applications are wide ranging from military and industrial to recreational, but its place of interest is nowhere but nature and inspiration from the way of migration of birds and hunting of archer fish. The pursuit problem involves one or more pursuers trying to catch a target that is moving in a certain direction. In this article, we delve into two modes of movement: movement on a straight line and movement on a curve. Our primary focus is on the latter. Within the context of movement on a straight line, we explore two methods and compare their respective results. Furthermore, we investigate the movement of two particles chasing each other and extend these findings to N particles that are chasing each other in pairs. By leveraging these two modes of movement, we present a novel relationship for two-particle and N-particle systems in pursuit. Lastly, we analyze the movement of moths around a lamp and evaluate their motion in relation to two-particle and N-particle systems in pursuit. The results of this analysis are carefully examined. 
\end{abstract}

\maketitle

\section{Introduction}
\subsection{History}
The pursuit problem was apparently first stated by Leonardo da Vinci. In this problem, the cat is chasing the mouse while the speed of both of them is constant and the mouse is moving in a straight line \cite{guha1994leonardo}.In 1732, a problem called (bugs), which was known as the dog, mouse, ant or beetles problem, was investigated by the French scientist Pierre Bouguer. These problems originate from the mathematics of pursuit curves\cite{marshall2003pursuit}. In 1877, Edward Lucas posed the three dog problem. It is stated in this problem: The dogs are first placed at the three vertices of the equilateral triangle, then each one follows the other. What is the curve that describes their motion? Three years later, Henri Brocard showed that the dog's path is a logarithmic spiral and the dogs eventually meet at a common point, this common point is called Brocard point of a triangle\cite{marshall2003pursuit}. For N bugs that are initially located on the vertices of regular N-polygon and follow each other at a constant speed, their trajectory is a logarithmic spiral and finally they all meet at the center of the polygon\cite{bernhart1959polygons}.
\subsection{Applications}

It has been suggested in \cite{kaden2022maneuvers} that air transportation will reduce its emissions in the near future. One proposed method to achieve this is by using the upwash of the wake vortex generated by a leading aircraft. This concept is inspired by migratory birds; however, the main challenge is overcoming the difficulties associated with pursuing the leading aircraft, which can be resolved by implementing a generalized solution of the chase problem. Multiple unmanned aerial vehicles (UAVs) or mobile robots, as discussed in \cite{hattenberger2007autonomous,galzi2006uav}, require a generalized solution of the chase problem, similar to that discussed in \cite{kaden2022maneuvers}.

\begin{figure}[h]
\includegraphics[width=40mm]{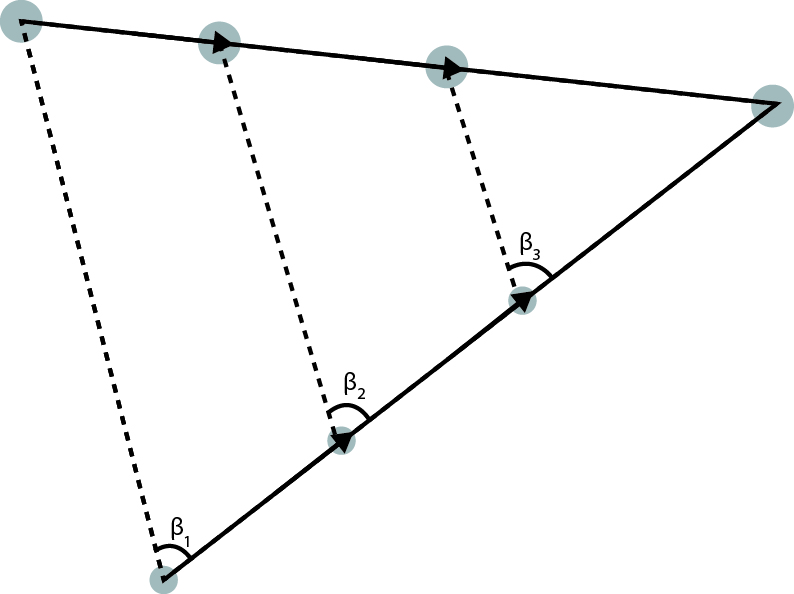}
\caption{\label{fig:enter-labe} New strategy}
\end{figure}

There is a wide range of applications for controlling multiple UAVs, such as exploration or surveillance, military or civilian scenarios, as noted in \cite{hattenberger2007autonomous,mungan2005classic}. Additionally, \cite{shaffer2009football}describes a new strategy, called the interception strategy, for American football players to catch the ball carrier. This strategy was commonly used when the sky was overcast shown in FIG\cite{shaffer2009football}.~\ref{fig:enter-labe}.

Houseflies, like migratory birds, tend to follow the leader and choose their path based on the angula velocity of the chasing fly \cite{mungan2005classic,land1974chasing}. The cinematography of the chase problem was used to measure the path taken by teleost fish and archer fish when catching their food \cite{lanchester1975pursuit,rossel2002predicting,wohl2006hunting}. Furthermore, the initial angle and distance strongly influence the route selected by teleost fish and archer fish to avoid collisions or choose the best path. This was noted in \cite{fajen2003behavioral}. Lastly, cinematography is required to target a quarry when only a partial location is known, as well as to control impact direction or limit the minimum radius curve (which maximizes acceleration) \cite{mungan2005classic}. A framework for controlling the formation of a group of robots was developed, and one critical parameter for this framework is the leader-follower graph \cite{wang1991navigation,das2002vision,sattigeri2003adaptive}.

\section{STRAIGHT LINE CHASING PROBLEM}
For many years, the issue of chasing with a straight line has been of great importance and has been extensively studied. This problem involves examining two main types of straight lines: those that are parallel to the horizontal or vertical axis, and those that are angled with respect to these axes.

\begin{figure}[h]
    \centering
    \includegraphics{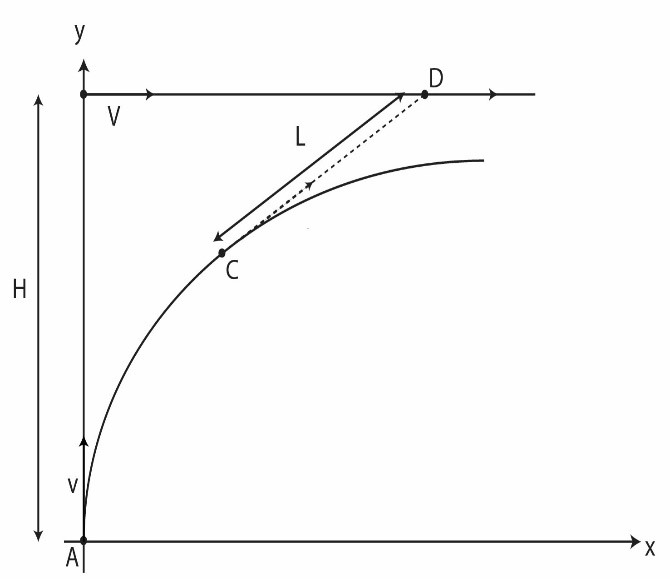}
    \caption{First type} 
    \label{Picture2.jpg}
\end{figure}

The first type of straight line path is a special case of the second, and both types are crucial to understanding this problem and finding potential solutions. To begin solving this problem, we can utilize the equation of the curve's slope for the first type of straight line path (as shown in FIG.~\ref{Picture2.jpg}). This approach was presented in \cite{mungan2005classic} and can be followed by proceeding with the rest of the necessary work.
\begin{figure}[h]
    \centering
    \includegraphics{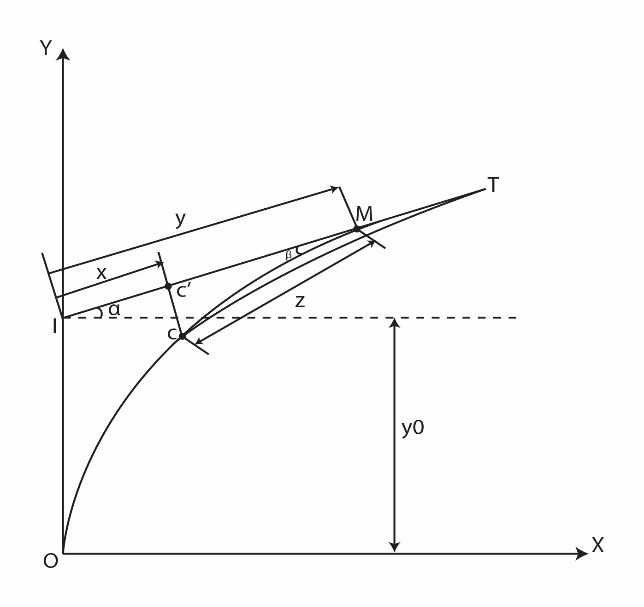}
    \caption{Second type} 
    \label{fig:artemis-logo}
\end{figure}
\begin{equation}
V t - x = (H- y)\frac{v_x}{v_y}\\
\end{equation}
By adopting this approach, they were able to derive various equations, including the equation for the distance between two objects in motion at a given moment, the equation for the vertical component of velocity, and the pursuit curve.Additionally, the equations can be used to determine the collision point.

\begin{equation}
    z = 1 - \frac{y}{H}
\end{equation}
\begin{equation}
    v_y =  \frac{2v}{z^{\frac{-V}{v}}+z^\frac{V}{v}}
\end{equation}

\begin{equation}
    L = \frac{H}{2}(z^{1-\frac{V}{v}}+z^{1+\frac{V}{v}})
\end{equation}
\begin{equation}
    \frac{x}{H} = (\frac{z^2-1}{4}) - \ln(\sqrt{z})
\end{equation}
As discussed in \cite{mungan2005classic}, the problem was solved in \cite{Palmaccio} by defining the slope (follower). However, the difference this time was that the slope was expressed as the derivative of y with respect to x. Using method \cite{Palmaccio}, it is also possible to identify the collision location. Essentially, the fundamental concepts were the same in both methods, and the solution could be obtained.

The two methods discussed above have been utilized to resolve the problem. Another approach is presented in \cite{chashchina2007dog}. In this method, rather than examining the slopes, location vectors are initially assigned to both movers. Since the pursuer is constantly moving towards the pursued, this yields the following equation:
\begin{equation}
    \Vec{r}_1 - \Vec{r}_2 = k(t)\Vec{V}
\end{equation}
where:

\begin{equation}
k(0) = \frac{L}{V}
\end{equation}
\begin{equation}
    k(T) = 0
\end{equation}
By utilizing the equations mentioned above, it is possible to determine the arrival time of both movers, as well as the pursuit curve (which was also obtained in \cite{Palmaccio}). Additional information regarding this can be found in \cite{chashchina2007dog}.

We will now explore the techniques for solving the second type of problems. In \cite{guha1994leonardo}, one of the methods to resolve this problem is presented. The scenario involves a cat chasing a mouse that moves along a straight line, making an angle of $\alpha$ with the horizon line. The cat moves at a speed of kv (where k > 1), while the mouse moves at a speed of v. M is the intersection point of the tangent line at point C on the line IT, creating the angle $\beta$ (FIG.~\ref{fig:artemis-logo}).
        
The solution to the problem can be summarized by the following three key equations:

\begin{equation}
    \frac{dx}{dt} = kvcos\beta
\end{equation}
\begin{equation}
    \frac{dy}{dt} = v
\end{equation}
\begin{equation}
    \frac{dz}{dt} = vcos\beta - kv
\end{equation}
As evident, the solution approach in this case is comparable to the previous two methods of the first type, except for the fact that the angle's impact needs to be taken into account since the path is not parallel to the horizon line. The problem is ultimately resolved by utilizing the aforementioned equations. For instance, the arrival point of the two mobiles can be determined as follows:
\begin{equation}
    X_T = y_0\frac{cos\alpha(k+sin\alpha)}{k^2-1}
\end{equation}
\begin{equation}
    Y_T = y_0(1+\frac{k+sin\alpha}{k^2-1}sin\alpha)
\end{equation}
Moreover, this method derives the pursuit differential equation (refer to \cite{guha1994leonardo} for more information).

Next, we will examine an alternative method introduced in \cite{barton2000pursuit}. The approaches outlined in \cite{barton2000pursuit} are considered general methods for resolving all pursuit problems, even if the motion functions are non-linear and intricate. The following equations govern the pursuer and pursued coordinate systems, where (x, y) represents the former and (X, Y) represents the latter:
\begin{equation}
    X = x + \lambda\dot x
\end{equation}
\begin{equation}
    Y = y + \lambda\dot y
\end{equation}
\begin{equation}
    \dot x^2 + \dot y^2 = c^2(\dot X^2 + \dot Y^2)
\end{equation}

The equations are rewritten below, taking into account the angle $\alpha$ formed with reference to the horizontal axis:
\begin{equation}
    x + \lambda\dot x = t cos\alpha
\end{equation}
\begin{equation}
    y + \lambda\dot y = y_0 + t sin\alpha
\end{equation}
\begin{equation}
    \dot x^2 + \dot y^2 = c^2
\end{equation}

The collision position determined from the equations is in agreement with the collision location stated in method \cite{guha1994leonardo}. It can be asserted that among all these techniques, \cite{barton2000pursuit} is the most efficient. This is because it offers a general principle and can analyze more intricate movements with greater ease.

\section{ROTATIONAL MOVEMENT}
\subsection{What is the problem?}
Consider a scenario where a fox with a speed of v is chasing a rabbit with the same speed. The rabbit moves at a path where its velocity makes an angle of $\alpha$ with respect to the line connecting the two entities. If their initial distance is l, where and when will they meet?\cite{Montazeri}
\subsection{Why this problem?}
The determination of the rabbit and fox's arrival time and position is not a primary focus of this article. The solution can be found in the same book \cite{Montazeri} and in similar studies \cite{chashchina2007dog,barton2000pursuit,Irodov,ptak1996dog}. What is significant in this context is the behavior and movement pattern of the hunter and prey in such systems. Furthermore, the study of this behavior in complex systems is of great importance \cite{marshall2003pursuit,marshall2004formations}. Most of the previous investigations \cite{chashchina2007dog,ptak1996dog}have solely considered the convergence point of the two entities.
\subsection{Solution}
\begin{figure}[h]
    \centering
    \includegraphics{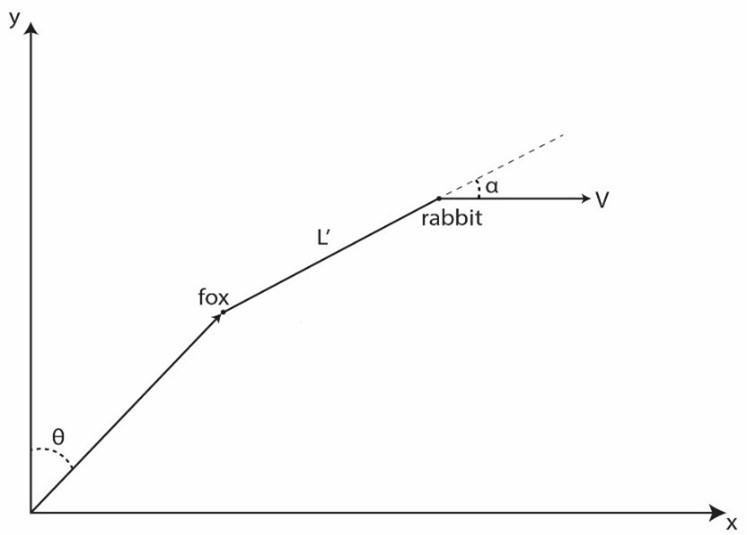}
    \caption{Motion’s diagram of fox and rabbit in Cartesian coordinate system As we see rabbit's velocity makes an angle of $\alpha$ with respect to the line connecting the two entities(L$^\prime$).}
\end{figure}
\begin{equation}
    \label{eq:e.20}
    \dot \theta = \frac{d\theta}{dt}=\frac{vsin\alpha}{L-(1-v)cos\alpha}
\end{equation}
\begin{equation}
    \label{eq:e.21}
    \bigtriangleup v (\hat{v}) = v cosd\theta - v = 0
\end{equation}
\begin{equation}
     \label{eq:e.22}
     \lim_{\bigtriangleup t\to 0}\bigtriangleup v(\hat{\theta}) = dv(\hat{\theta}) = v sin d\theta = vd\theta
\end{equation}
\begin{equation}
    \label{eq:e.23}
    dv = \sqrt{dv(\hat{\theta})+dv(\hat{v})} \xRightarrow{dv(\hat{v}) = 0} dv = dv(\hat{\theta})
\end{equation}
Using Eqs.~(\ref{eq:e.20}), (\ref{eq:e.21}), (\ref{eq:e.22}) and (\ref{eq:e.23}) then we can write:
\begin{equation}
    L' = L- (1-v)t \cos{\alpha}    
\end{equation}
\begin{equation}
\begin{split}
    \label{eq:e.25}
    d\theta = (\frac{v \sin{\alpha}}{L-(1-v)t \cos{\alpha}})dt \\
    \xRightarrow{dv=vd\theta} dv = (\frac{v \sin{\alpha}}{L-(1-v)t\cos{\alpha}})vdt
\end{split}
\end{equation}
\begin{equation}
    \label{eq:e.26}
    dv(\hat{x}) = dv \cos{\theta}
\end{equation}
\begin{equation}
    \label{eq:e.27}
    dv(\hat{y}) = -dv \sin{\theta}
\end{equation}
\begin{equation}
    \label{eq:e.28}
    \sin{\theta}=\frac{dx}{\sqrt{(dx^2+dy^2)}}\xRightarrow{\frac{1}{dt}} \sin{\theta} = \frac{\frac{dx}{dt}}{(\frac{\sqrt{(dx^2+dy^2)}}{dt})} = \frac{v(x)}{v}
\end{equation}
\begin{equation}
    \label{eq:e.29}
    \cos{\theta}=\frac{dy}{\sqrt{(dx^2+dy^2)}}\xRightarrow{\frac{1}{dt}} \cos{\theta} = \frac{\frac{dy}{dt}}{(\frac{\sqrt{(dx^2+dy^2)}}{dt})} = \frac{v(y)}{v}
\end{equation}
By merging Eqs.~(\ref{eq:e.25}), (\ref{eq:e.26}), (\ref{eq:e.27}), (\ref{eq:e.28}), and (\ref{eq:e.29}), it is feasible to formulate two coupled differential equations for velocity in two directions.
\begin{equation}
    \label{eq:e.30}
    dv(\hat{x}) = (\frac{vv(\hat{y})\sin{\alpha}}{L-v(1-\cos{\alpha})t})dt
\end{equation}
\begin{equation}
    \label{eq:e.31}
    dv(\hat{y}) = -(\frac{vv(\hat{x})\sin{\alpha}}{L-v(1-\cos{\alpha})t})dt
\end{equation}
\section{RESULT}
The approach we have employed thus far is somewhat similar to the method presented in \cite{Montazeri} for determining the endpoint and we can determine the end point without solving Eqs.~(\ref{eq:e.30}) and (\ref{eq:e.31}). However, our primary focus is to examine the motion parameters, such as velocity and location, along the direction of motion. To achieve this, we can solve the coupled differential equations of the system using software tools like Mathematica.
Solving with Mathematica:
\begin{equation}
\begin{split}
    v_x(t) = C_1\cos{[\frac{\sin{\alpha}\ln{(L+(-1+\cos{\alpha})tV)}}{-1+ \cos{\alpha}}]} + \\
    C_2\sin{[\frac{\sin{\alpha}\ln{(L+(-1+\cos{\alpha})tV)}}{-1+\cos{\alpha}}]}
\end{split}
\end{equation}

\begin{equation}
\begin{split}
    v_y(t) = C_2\cos{[\frac{\sin{\alpha}\ln{(L+(-1+\cos{\alpha})tV)}}{-1+ \cos{\alpha}}]} - \\
    C_1\sin {[\frac{\sin{\alpha}\ln{(L+(-1+\cos{\alpha})tV)}}{-1+\cos{\alpha}}]}
\end{split}
\end{equation}
Well, we can take the integral of these equations to give us the coordinate equation for motion over time:
\begin{multline}
    x(t) =  B_1 +
    {(L+(-1+\cos{\alpha})tV)} \times \\ 
    (\frac{((-1+\cos{\alpha})C_1-\sin{\alpha}C_2)\cos{[\frac{\sin{\alpha}\ln{[L+(-1+\cos{\alpha)tV}]}}{-1+\cos{\alpha}}]}}{{(1-2\cos{\alpha}+\cos{\alpha}^2+\sin{\alpha}^2)V}}+\\ \frac{(\sin{\alpha}C_1+(-1+\cos{\alpha})C_2)\sin{[\frac{\sin{\alpha}\ln{[L+(-1+\cos{\alpha})tV]}}{-1+\cos{\alpha}}]}}{{(1-2\cos{\alpha}+\cos{\alpha}^2+\sin{\alpha}^2)V}}) \\
\end{multline}
\begin{multline}
        {y(t) = B_2 + {(L+(-1+\cos{\alpha})tV)((\sin{\alpha}C_1+(-1+\cos{\alpha})}} \times\\ \frac{C_2\cos{[\frac{\sin{\alpha}\ln{[L+(-1+\cos{\alpha)tV}]}}{-1+\cos{\alpha}}]}}{(1-2\cos{\alpha}+\cos{\alpha}^2+\sin{\alpha}^2)V}+\\ \frac{(C_1-\cos{\alpha}C_1+\sin{\alpha}C_2)\sin{[\frac{\sin{\alpha}\ln{[L+(-1+\cos{\alpha})tV]}}{-1+\cos{\alpha}}]}}{(1-2\cos{\alpha}+\cos{\alpha}^2+\sin{\alpha}^2)V})
\end{multline}
Note that the coefficients $C_1$,$C_2$,$B_1$  and $B_2$ are established based on the initial conditions. We have flexibility in selecting the coordinates by rotating them, and we opt for the coordinates that satisfy the following initial conditions:
\begin{multline}
    v_x (0)=0,v_y (0)=v,x(0)=0,y(0)=0 \\
    v_x(t) = \frac{-v\cos{[\frac{[\sin{\alpha}\ln{[L+(-1+\cos{\alpha})tv]}}{-1+\cos{\alpha}}]}\sin{[\frac{\sin{\alpha}\ln{[L]}}{-1+\cos{\alpha}}]}}{\cos{[\frac{\sin{\alpha}\ln{[L]}}{-1+\cos{\alpha}}]^2}+\sin{[\frac{\sin{\alpha}\ln{[L]}}{-1+\cos{\alpha}}]^2}} +  \\ \frac{v\cos{[\frac{\sin{\alpha}\ln{[L]}}{-1+\cos{\alpha}}]}\sin{[\frac{\sin{\alpha}\ln{[L+(-1+\cos{\alpha})tV]}}{-1+\cos{\alpha}}]}}{\cos{[\frac{\sin{\alpha}\ln{[L]}}{-1+\cos{\alpha}}]^2}+\sin{[\frac{\sin{\alpha}\ln{[L]}}{-1+\cos{\alpha}}]^2}}
\end{multline}

\begin{multline}
    v_y(t) = \frac{(v\cos{[\frac{\sin{\alpha}\ln{[L]}}{-1+\cos{\alpha}}]}\cos{[\frac{\sin{\alpha}\ln{[L+(-1+\cos{\alpha})tv]}}{-1+\cos{\alpha}}]}}{(\cos{[\frac{\sin{\alpha}\ln{[L]}}{-1+\cos{\alpha}}]^2}+\sin{[\frac{\sin{\alpha}\ln{[L]}}{-1+\cos{\alpha}}]^2})}+\\
    \frac{v\sin{[\frac{\sin{\alpha}\ln{[L]}}{-1+\cos{\alpha}}]}\sin{[\frac{\sin{\alpha}\ln{[L+(-1+\cos{\alpha})]tv}}{-1+\cos{\alpha}}]})}{(\cos{[\frac{\sin{\alpha}\ln{[L]}}{-1+\cos{\alpha}}]^2}+\sin{[\frac{\sin{\alpha}\ln{[L]}}{-1+\cos{\alpha}}]^2})}
\end{multline}
We set the initial conditions as follows (note that the specific choice of initial conditions does not impact the overall outcome of our movement analysis):

v=2 (m/s)  ,L=20(m), $\alpha$ = $\pi$/3

\begin{figure}[h]
    \includegraphics[width=45mm]{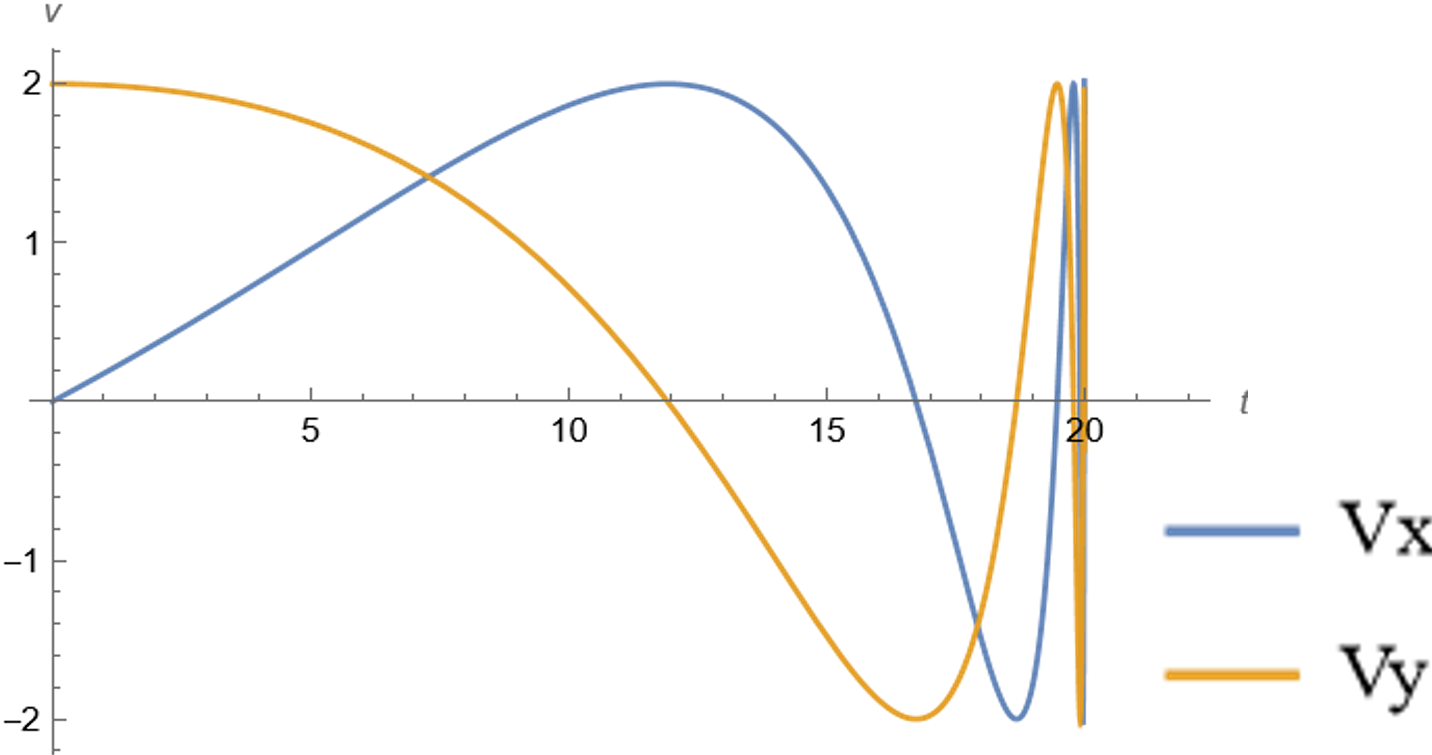}
    \caption{graph of the Fox's speed over time} 
      \label{fig:m.5}
\end{figure}
\begin{figure}[h]
    \centering
    \includegraphics[width=45mm]{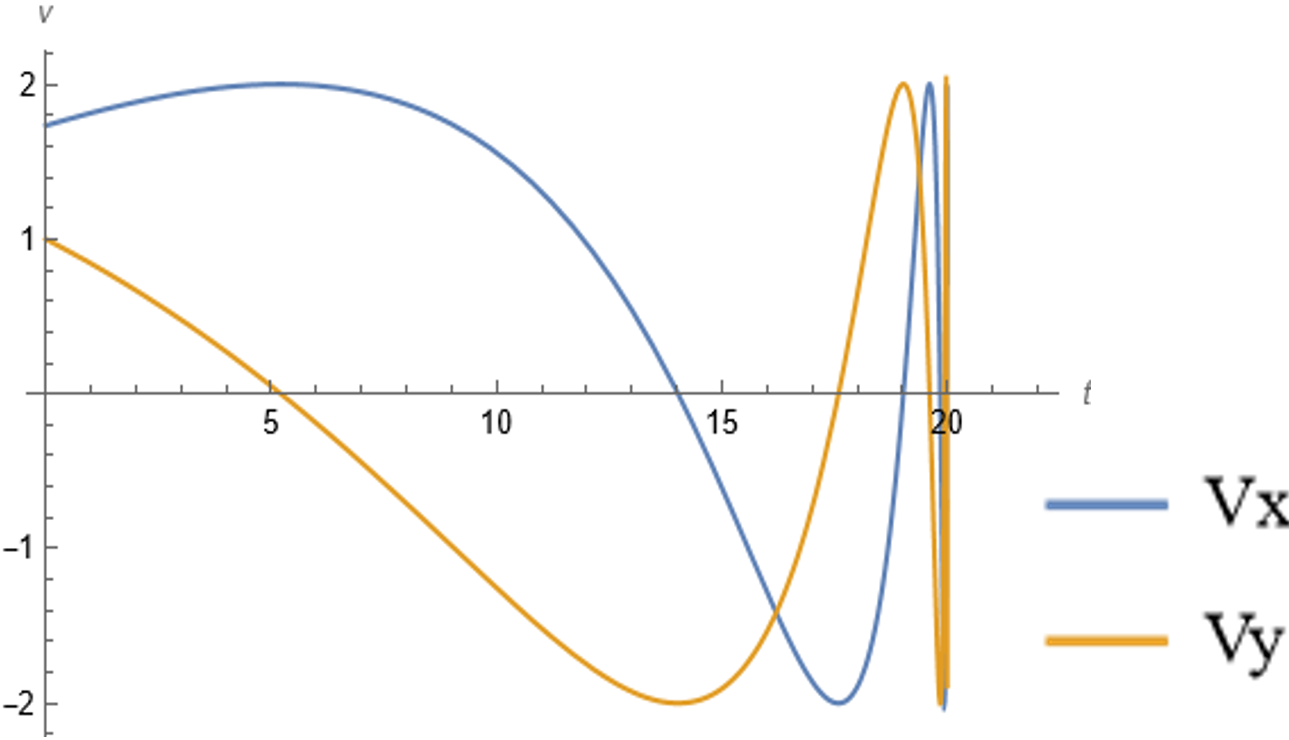}
    \caption{A graph of the rabbit's speed over time} 
      \label{fig:m.6}
\end{figure}

FIG.~\ref{fig:m.5} and FIG.~\ref{fig:m.6} depict the initial phenomenon that occurs in this system, where the changes in speed begin to increase with the passage of time. In the twentieth second, these changes, including acceleration and a sharp increase in speed with a sign change, become more prominent. To analyze this event, we need to examine the system in a relative device. In a relativistic device, the length of the connecting line decreases (radius of gyration), leading to an increase in angular velocity changes. At the point where the two moving entities meet, this length tends to zero, causing our rotational acceleration to tend towards infinity. Furthermore, the time required to complete a round tends to zero, resulting in an infinite acceleration over time, and an increase in the speed of changing the speed sign due to the reduction in rotation time. (the two moving parts rotate together)

It is worth noting that we have formulated the equations of motion for the fox, and the equations of motion for the rabbit are identical to those of the fox, except for the boundary conditions.
\begin{figure}[h]
    \includegraphics[width=80mm]{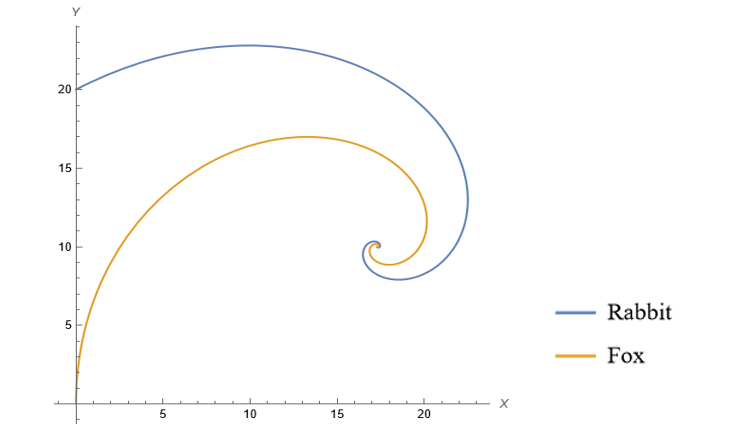}
    \caption{Fox and rabbit movement path diagram in Cartesian coordinates} 
      \label{fig:m.7}
\end{figure}
FIG.~\ref{fig:m.7}, we observe that the shape of the rabbit's movement path is similar to that of the fox, except that the initial conditions in the equations show that we have shifted the fox's diagram by the initial distance L and rotated it to the right by $\alpha$. This indicates that we can convert the fox diagram into a rabbit diagram, as mentioned earlier.
\begin{figure}[h]
    \centering
    \includegraphics{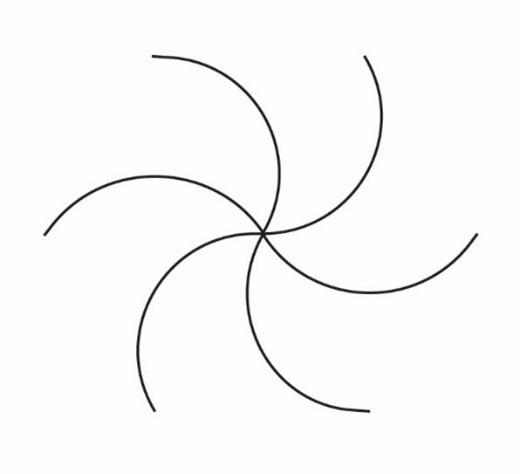}
    \caption{When the $\protect\alpha$ = 60$^{\circ}$ the initial conditions of our two-particle system exhibit behavior that is similar to that of a six-particle system. Additionally, this behavior displays the same symmetry as the figure opposite to it.}
  \label{fig:m.8}
\end{figure}

Another conclusion that we can draw from FIG.~\ref{fig:m.7} is that the rabbit moves like a fox that is chasing a rabbit in the direction of its initial speed, while being located at a distance of L. If we continue this reasoning, we can assume that the rabbit behaves like a fox that is pursuing another rabbit. Thus, the system of two particles following each other behaves like a system of N particles following each other (FIG.~\ref{fig:m.8}).

For situations where $\alpha$ is a divisor of 360, our N-particle system becomes a closed system, where the N bodies follow the first body due to the constraint of movement in the connecting line. The interesting thing is that this N-particle system behaves like a two-particle system. Therefore, we can use the resulting relationship for an N-particle system for a two-particle system.

For a closed N-particle system, we know that the shape symmetry system preserves its initial shape. This symmetry implies that we can write the rates of motion in the polar device to the center of this closed system.As the system is symmetrical, the point of convergence of the particles is certainly at its center.
\begin{figure}[h]
    \centering
    \includegraphics{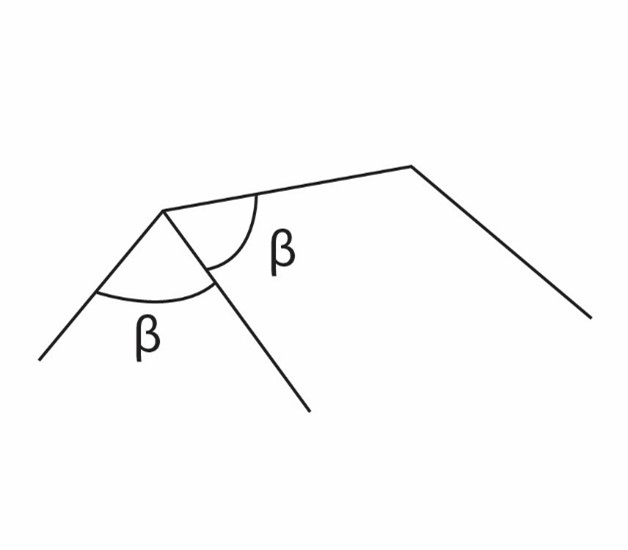}
    \caption{In a generalized two-particle to N-particle system, the angle between the center of rotation and each vertex of the N polygon that is con-structed is equal to beta. This relationship holds true regardless of the number of parti-cles involved in the system, allowing for a gen-eralized approach to the problem.}
\end{figure}
\begin{figure}
    \centering
    \includegraphics{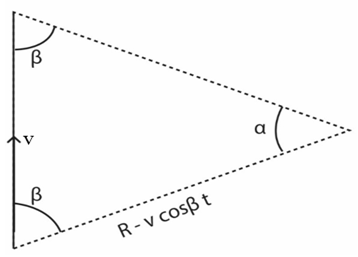}
    \caption{we can observe the triangle that is formed by connecting $N_i$,$N_{i+1}$ and the center of rotation. By analyzing the motion of the particle in two directions of polar coordinates. We can generilized this method for every paticle.}
\end{figure}
\begin{equation}
    \label{eq:e.38}
    2 \beta = \pi - \alpha \xRightarrow{\frac{1}{2}} \beta = \frac{\pi}{2} - \frac{\alpha}{2}
\end{equation}
\begin{equation}
    \label{eq:e.39}
    2R\sin{\frac{\alpha}{2}} = l \xrightarrow{} R = \frac{l}{2\sin{\frac{\alpha}{2}}}
\end{equation}

\begin{equation}
    \label{eq:e.40}
    v = \sqrt{(\dot{r})^2+(r\dot{\theta})^2}
\end{equation}
\begin{equation}
    \label{eq:e.41}
    \dot{r}  = -v \cos{\beta} = -v \sin{\frac{\alpha}{2}}
\end{equation}

\begin{equation}
    \label{eq:e.42}
    r = R - v \sin{\frac{\alpha}{2}}t \xrightarrow{} r = \frac{l}{2\sin{\frac{\alpha}{2}}} - vt \sin{\frac{\alpha}{2}}
\end{equation}
Using Eqs.~(\ref{eq:e.38}), (\ref{eq:e.39}), (\ref{eq:e.40}),(\ref{eq:e.41}) and (\ref{eq:e.42}) then we can write:
\begin{equation}
    v = \sqrt{(-v\sin{\frac{\alpha}{2}})^2+((\frac{l}{2\sin{\frac{\alpha}{2}}}-v\sin{\frac{\alpha}{2}}t)\dot{\theta})^2}
\end{equation}

\begin{equation}
    \dot{\theta} = \frac{v \cos{\frac{\alpha}{2}}}{(\frac{l}{2\sin{\frac{\alpha}{2}}}- v \sin{\frac{\alpha}{2}}t)}
\end{equation}

If we solve this differential equation
\begin{equation}
    \label{eq:e.45}
    - \frac{\theta}{\cot{\frac{\alpha}{2}}} = \ln{(\frac{l}{2\sin{\frac{\alpha}{2}}}-v \sin{\frac{\alpha}{2}}t)} + c
\end{equation}

 c is a constant determined by the initial conditions

Using Eqs.~(\ref{eq:e.42}) and (\ref{eq:e.45}) we can write :
\begin{equation}
    \label{eq:e.46}
    r(\theta) = r_0 e^{-\frac{\theta}{\cot{\frac{\alpha}{2}}}}
\end{equation}
Eq.~(\ref{eq:e.46}) is applicable to all the systems we have analyzed so far, including the two-particle system in pursuit and N-particle systems' motion behavior in polar coordinates.
Eq.~(\ref{eq:e.46}) can also be observed in moth insects. In the past, when the air was dark due to the absence of lamps and lighting in modern cities, moths used the moon for orientation and movement, keeping the angle between the direction of their movement and the distance vector between the moth and the moon constant. This was possible because the moon was considered to be at an infinite distance\cite{WinNT}. However, with the increase in brightness in cities and the prevalence of lamps, the light from lamps became brighter than that of the moon, causing moths to choose lamps as a source of movement. By approaching the lamp, the moth tries to keep the angle between the vector connecting to the light source and the direction of movement constant, resulting in a rotational movement similar to the system of two tracked particles. It can be argued that the motion of the two-particle system in pursuit and the N-particle system in pursuit, previously investigated, moves with a constant angle between the line connecting the particle to the center of rotation and the velocity vector. This movement is similar to the movement of moths.

\bibliographystyle{unsrt}
\bibliography{aipsamp}
\end{document}